\documentstyle[preprint,aps,graphicx,rotating]{revtex}

\tightenlines
\begin{document}

\title{Assessment of exchange-correlation functionals for the calculation
of dynamical properties of small clusters in TDDFT}
\author{M.~A.~L.~Marques\(^{*\dagger}\) \and Alberto~Castro\(^*\) and Angel~Rubio\(^\ddagger\)\footnote{
On sabbatical leave from Departamento de F\'{\i}sica Te\'orica, 
Universidad de Valladolid, E-47011 Valladolid, Spain
}}
\address{\(^*\)Departamento de F\'{\i}sica Te\'{o}rica, 
Universidad de Valladolid, E-47011 Valladolid, Spain}
\address{\(^\dagger\)Centro de F\'{\i}sica Computacional,
Departamento de F\'{\i}sica da Universidade de Coimbra,
Rua Larga, 3004-516, Coimbra, Portugal}
\address{\(^\ddagger\)
Laboratoire de Solides Irradi\'{e}s (LSI), CNRS-CEA, \'Ecole Polytechnique, F-91128 Palaiseau,
France and \\
Donostia International Physics Center (DIPC), 20018 San Sebastian (Spain)}
\date{\today}
\maketitle
\begin{abstract}
We present a detailed study of different exchange-correlation (xc) functionals
in describing the dynamical properties of finite systems. For that purpose,
we calculated the static polarizabilities, ionization potentials and
optical absorption spectrum of four small clusters,
Na\(_2\), Na\(_4\), SiH\(_4\) and Si\(_2\)H\(_6\), using a real-space,
real-time technique.
The computed static polarizabilities and ionization potentials 
seem to be in rather good agreement with the available experimental data,
once the proper asymptotics of the potential are taken into account.
The same conclusion holds for the absorption spectra, although
the xc kernels in use do not provide a
sufficiently strong attractive interaction between electrons and holes, leading to
spectra slightly shifted towards higher energies. This deficiency is traced 
back to the insufficient description of dynamical effects in the correlation functional.
Furthermore, it is shown that the xc potential
used to obtain the ground state is the key factor
to get reasonable spectra, whereas the choice of the xc kernel just
amounts to small, although important, quantitative changes.
\end{abstract}

\newcommand{\be}{\begin{equation}}
\newcommand{\ee}{\end{equation}}
\newcommand{\bea}{\begin{eqnarray}}
\newcommand{\eea}{\end{eqnarray}}
\newcommand{\br}{{\bf r}}

\section{Introduction}
Density Functional Theory (DFT)\cite{bDFT} became, in the two last decades,
the method of election for the {\it ab initio} calculation of
material properties. Systems with a couple thousand atoms are
now routinely investigated, and the calculated energies,
geometries, etc., very often agree spectacularly with the experimental
data. This success, however, was not immediate: More than twenty
years of research were necessary to obtain an exchange-correlation (xc)
energy functional which was precise enough to satisfy the 
quantum-chemistry needs. The accurate and sophisticated 
Generalized Gradient Approximations (GGA) that we now have
at our disposal are indeed the result of a long history of attempts, 
tests and failures.

Despite these remarkable achievements,
there are some quantities which are beyond the reach of the 
conventional DFT theory. In this Article we will be concerned by one
of such properties, namely electronic excitation energies and optical spectra.
Several extensions of the original framework were put forward 
to obtain excited-state properties. From all different
approaches, Time-Dependent DFT (TDDFT)\cite{bTDDFT} emerged the last years as the
main DFT formalism for these calculations.
One of the main advantages of the TDDFT formulation is that it allows one
to deal with problems beyond the perturbative regime as, for example,
the response of atoms to ultra-short and intense laser beams. In
this way, by solving the time-dependent Kohn-Sham (KS) equations, we can get,
{\it e.g.},
information concerning the harmonic spectrum or the ionization yields.
This type of approach is very important in condensed matter
science where femtosecond laser pulses are used to monitor the dynamics
of electrons in a solid. Furthermore, with larger computational
power, we may deal with non-adiabatic couplings between the
electronic and the ionic degrees of freedom in presence of these
high electromagnetic fields (we may observe phonon assisted
structural transformation induced by these external fields).

Excitation energies
can be obtained from TDDFT either from the position of the poles of
the KS linear response function \cite{bExcitPole,bPeteretal},
or from the time-dependent density. In the second case, we will be using an
approximated xc time-dependent potential, \(v_{\rm xc}(\br, t)\),
while in the former, the key quantity is the xc kernel, 
\(f_{\rm xc}(\br,\br',t,t')\), defined by the functional derivative:
\be
        f_{\rm xc}(\br,\br',t,t') = \frac{\delta v_{\rm xc}(\br, t)}
        {\delta n(\br',t')}.
\ee
As expected, the xc kernel, due to its complicated structure,
is much harder to model than the local time-dependent xc potential.
Furthermore, both calculations usually start from the KS
ground-state of the system, which depends on the approximation
used for the static \(v_{\rm xc}(\br)\).

One of the most widely used approximation to time-dependent phenomena is
the adiabatic local density approximation (ALDA). Although this functional
is constructed using the xc correlation energy of the (constant density)
electron gas, it yields, like the LDA does in the static case, rather
accurate results for systems with rapidly varying densities, as
atoms, surfaces or clusters \cite{bALDA,bZang,bprlsi,bChelikowsky}. For example,
the photo-absorption spectrum of rare-atom gases was computed in 
Ref. \cite{bZang} and the agreement with experiment is remarkably
good. Results of similar quality have been achieved for the
photo-response of small metallic and semiconducting clusters
\cite{bprlsi}. For the metallic clusters it was shown that the inclusion
of xc effects in the dielectric response was important to get the
correct red-shift, as compared to experiments, while for the
semiconducting (silicon) clusters it was found that the spectra
of different isomers was sufficiently different to distinguish between
them. Surface and confinement effects were responsible for the
appearance of absorption in the optical range for the two silicon clusters,
Si\(_4\) and Si\(_6\).
Similar results were found for C\(_{20}\) clusters \cite{bacmalmar}.

The purpose of this work is twofold: To address the impact of a good
xc-potential in the optical spectrum of small clusters and to estimate
how relevant the \(f_{\rm xc}\) kernel is in this respect. We note that
in Ref.\cite{bPeteretal} the quality of different functionals was tested
for light atoms, and in Ref.\cite{Lein} for the calculation of the correlation energy
of the homogeneous electron gas.
In the first of these works, which included calculations using
the exact static xc potential, the correct 
description of the xc-potential used to obtain the ground state
seemed to be the dominant factor, whereas
the kernel played a marginal role. However this was not the case when looking
at the total correlation energy of the homogeneous electron gas.
In particular, the nonzero spatial
range of $f_{xc}(r,r',\omega)$ could not be neglected, whereas the
frequency dependence appeared to be less important. 
Our work clearly demonstrates that the optical spectrum of
small systems is not only determined by the static xc potential,
but also the kernel plays a role.

The rest of the Article is structured in the following way:
In Section \ref{StaticXC}, we give a brief overview of the
xc functional zoo used in static KS calculations; We then
proceed by explaining how these functionals can be changed in order 
to be used in TDDFT calculations; In Section \ref{MethodResults} we give some
details on the techniques used in our calculations, and present our results
for some selected sodium and silicon clusters; 
We finally conclude and give some remarks on the quality of 
the functionals tested. 

\section{XC Functionals for Static DFT}
\label{StaticXC}
The first (and simplest) approximation for the xc functional,
the Local Density Approximation (LDA) introduced by Kohn and
Sham in 1965 \cite{bKohnSham}, yielded remarkably good results for such a modest
effort, but the method didn't prove accurate enough to be used
in theoretical chemistry. The LDA xc energy functional can be written as
\be
        \label{LDA}
  E_{\rm xc}^{\rm LDA} = \int\!\! d^3 r\: n(\br)\: \epsilon^{\rm hom}_{\rm xc}(n(\br))
        ,
\ee
where \(n(\br)\) is the electronic density at point \(\br\), and
\(\epsilon^{\rm hom}_{\rm xc}(n)\) stands for the xc energy density
of the homogeneous electron gas with density \(n\). This function
is calculated very accurately using Monte-Carlo techniques
\cite{bCeperleyAlder}, and then fitted to some simple analytical 
function \cite{bPerdewZunger}. The functional (\ref{LDA}) has
several shortcomings, among which we point out the following:
\begin{itemize}
\item It neglects non-local effects, {\it i.e.} the LDA xc energy density
at point \(\br\) only depends on the density at that point. We therefore
should not expect this functional to work in cases where the density
has very strong spatial variations.
\item The exchange part of the functional does not cancel exactly the
self-energy part of the Hartree term. This leads to a wrong asymptotic
behavior of the xc potential for finite systems (it goes exponentially
to zero, instead as \(-e^2/r\)). Properties that strongly depend on this
asymptotic behavior, like the ionization potential of atoms and
molecules, come out with very large errors. Also,
there are no Rydberg states within the LDA, and negative ions usually
do not bind, which renders the calculation of electron-affinities impossible.
\item LDA usually overbinds, giving too short bond-lengths, etc.
\end{itemize}

The next generation
of functionals included the so-called Generalized Gradient Approximated (GGA) 
functionals \cite{bBecke,bLYP,bPBE}.
They can be written as:
\be
        \label{GGA}
  E_{\rm xc}^{\rm GGA} = \int\!\! d^3 r\: n(\br)\: \epsilon^{\rm GGA}_{\rm xc}
        (n(\br), \nabla n(\br))
        ,
\ee
where \(\nabla n(\br)\) is the gradient of the density at the point
\(\br\). \(\epsilon^{\rm GGA}_{\rm xc}\) is usually some analytic function
with some free parameters that are either fitted to experiment, or determined
by some exact sum-rules. GGAs solve some of the problems present in the
LDA, and in some cases yield results with high enough precision to be used
as a tool in quantum chemistry calculations.
Recently, a new class of functionals generalizing the GGAs
has been proposed. These so-called Meta-GGAs \cite{bMGGA}, depend 
explicitly not only on the density and its gradient, but also on the kinetic
energy density\footnote{
  The MGGAs, due to the explicit dependence on the kinetic energy density are,
in fact, not simple GGAs, but orbital functionals.
}.
This extra dependence adds more flexibility, and
allows better approximations to the exact xc functional to be built.
Unfortunately, all the GGAs and MGGAs proposed till now suffer, to different degrees, 
from the same self-interaction problem as the original LDA.
A way to circumvent it was proposed by van~Leeuwen and Baerends 
in 1994 \cite{bLB94}: They applied Becke's construction \cite{bBecke}
not to the derivation of the exchange energy functional
(like in Becke's original work), but to the modeling of the xc potential
directly. By imposing the correct asymptotic behavior to the potential, 
they were able to get much better ionization potentials (and eigenvalues
in general).

To complete the zoo of energy functionals, we still have to refer to orbital functionals,
in the so-called Optimized Effective Potential (OEP), or 
Optimized Potential Method (OPM) \cite{bEXX}.
This third generation of functionals is written explicitly in terms of the Kohn-Sham
orbitals (being nevertheless implicit functionals of the density):
\be
  E_{\rm xc}^{\rm OEP} = E_{\rm xc}^{\rm OEP}\left[\varphi_1(\br)\cdots\varphi_N(\br)\right]
  .
\ee
The xc potential is then calculated using twice the chain-rule for functional 
derivatives:
\bea
  \label{vOEP}
  \lefteqn{
    v_{\rm xc}^{\rm OEP}(\br) = \frac{\delta E^{\rm OEP}_{\rm xc}}{\delta n(\br)} 
        }\nonumber \\
  & = & \sum_{i=1}^N \int\!\! d^3 r'\:d^3 r''
        \left[\frac{\delta E^{\rm OEP}_{\rm xc}}{\delta \varphi_i(\br')}
        \frac{\delta \varphi_i(\br')}{\delta v_{\rm KS}(\br'')} + {\rm c.c.}
        \right] \frac{\delta v_{\rm KS}(\br'')}{\delta n(\br)}
  .
\eea
The first term of the right, 
\(\frac{\delta E^{\rm OEP}_{\rm xc}}{\delta \varphi_i(\br')}\),
can easily be obtained from the expression for 
\(E_{\rm xc}^{\rm OEP}\left[\varphi_1(\br)\cdots\varphi_N(\br)\right]\), while
the second, 
\(\frac{\delta \varphi_i(\br')}{\delta v_{\rm KS}(\br'')}\)
can be calculated using first-order perturbation theory. Finally, the
remaining functional derivative can be identified with the inverse of the response
function for non-interacting electrons, \(\chi^{-1}_{\rm KS}(\br,\br')\). Rearranging
the terms in Eq.~(\ref{vOEP}), we obtain an integral equation for
\(v_{\rm xc}^{\rm OEP}\). The solution  of this integral equation is 
numerically very involved, and has only been achieved for systems
with very high symmetry. An alternative is to perform an approximation
first proposed by Krieger, Lee and Iafrate\cite{bKLI} (KLI), which transforms the hard
task of solving an integral equation in three dimensional space
into the simple one of solving a small set of linear equations.

Two examples of these orbital functionals are the exact exchange
(EXX), and self-interaction corrected (SIC) LDA 
functionals \cite{bPerdewZunger}. In the
first, one employs the exact expression for the exchange energy:
\be
  \label{EXX}
        E_{\rm x}^{\rm EXX} = -\frac{1}{2} \sum_\sigma\sum_{i, j = 1}^N
        \int\!\! d^3 r\: d^3 r' \frac
        {\varphi^*_{j\sigma}(\br)\varphi^*_{i\sigma}(\br')
         \varphi_{i\sigma}(\br)\varphi_{j\sigma}(\br')}
        {\left|\br-\br'\right|}.
\ee
The SIC-LDA functional, originally proposed by Perdew and Zunger \cite{bPerdewZunger}, 
can be written as:
\bea
  \label{eqSIC}
  E_{\rm xc}^{\rm SIC} & = & 
        E_{\rm xc}^{\rm LDA}\left[n_\uparrow(\br),n_\downarrow(\br)\right] -
        \sum_\sigma\sum_{i= 1}^N E_{\rm xc}^{\rm LDA}
        \left[\left|\varphi_i(\br)\right|^2,0\right]
  \nonumber \\ & &
  - \frac{1}{2} \sum_\sigma\sum_{i= 1}^N
        \int\!\! d^3 r\: d^3 r' \frac
        {\left|\varphi_{i\sigma}(\br)\right|^2
         \left|\varphi_{i\sigma}(\br')\right|^2}
        {\left|\br-\br'\right|}.
\eea
It is clear that the SIC-LDA obeys two of the features of the exact
functional: It exactly cancels the self-interaction part of the
Hartree energy and it vanishes for one electron systems. The SIC-LDA functional is
nevertheless ill-defined, for it is not invariant upon an
unitary transformation of the Kohn-Sham wave-functions.

\section{XC Functionals for TDDFT}
\label{TDXC}
The simplest, and most commonly used, approximation to the 
xc functional in TDDFT is the adiabatic LDA (ALDA), in which
the static LDA functional is used for the dynamical properties,
but evaluated at the time-dependent density:
\be
  v_{\rm xc}^{\rm ALDA}(\br,t) = v_{\rm xc}^{\rm hom}(n(\br,t))
  .
\ee
In the ALDA, the \(f_{\rm xc}\) kernel is a contact function
in time and space:
\be
  f_{\rm xc}^{\rm ALDA}(\br,t;\br',t') = \delta(t-t') \delta(\br-\br')
  \left. \frac{d v_{\rm xc}^{\rm hom}(n)}{d n} \right|_{n=n(\br,t)}
  .
\ee
Following the same reasoning, it is straightforward to write 
adiabatic GGA potentials and \(f_{\rm xc}\) kernels. Unfortunately,
the thresholds for the onset of absorption calculated 
either with adiabatic LDA or GGA functionals
are typically below the observed ones (by several
eV in the case of atoms). This is an intrinsic drawback of the ALDA/GGA because,
in principle, TDDFT should yield the correct thresholds. This problem
is once more related to the wrong asymptotic behavior of the
effective Kohn-Sham potential that goes exponentially to zero
instead of \(-e^2/r\) for neutral systems. This is due,
as already mentioned in the previous Section, to the insufficient
correction of the self-interaction part of the Hartree potential.
Also the xc kernel, \(f_{\rm xc}\), suffers from a self-interaction
error. A simple way to correct the asymptotic part of the adiabatic
potential is using the Adiabatic LB94 approximation. However,
the high lying excitation energies calculated with this functional
are usually overestimated for small molecules, and the LB94 does not do quite
as well as the ALDA for the low-lying states \cite{bCasida}.

A generalization of the EXX potential to the time-domain is also
possible \cite{bTDOEP}. It starts from the perturbative expansion of the action
functional and then uses the chain-rule for functional derivatives
to derive an integral equation for the xc potential. A KLI-like
approximation can then be used to simplify the task of solving this
integral equation. Using the same procedure, one can derive an
integral equation for \(f_{\rm xc}\). However, for practical calculations,
it is desirable to devise a simple analytical expression for the xc kernel.
To this end, it is common practice to further simplify the KLI
potential by neglecting one of its terms \cite{bTDOEP}. In the static
exchange-only case, this leads to the so-called Slater
approximation \cite{bSlater}. \(f_{\rm xc}\) then reads:
\be
  f_{{\rm xc}\;\sigma \sigma'}^{\rm EXX\:approx.}(\br t, \br' t') =
  -\delta(t-t')\delta_{\sigma\sigma'} \frac
  {\left|\sum_{i = 1}^N \varphi_{i\sigma}(\br)\varphi^*_{i\sigma}(\br')\right|^2}
  {\left|\br-\br'\right|n_\sigma(\br)n_\sigma(\br')}
  .
\ee
One should note that, due to the
extra functional derivative in the definition of the xc kernel,
it is much more complicated, in this case, to evaluate directly
\(f_{\rm xc}\) than to time-propagate the KS states with \(v_{\rm xc}\).

Finally, an adiabatic SIC-LDA can be easily
written in the spirit of Eq. (\ref{eqSIC}).

Other functionals designed to retain features which are important
in time-dependent calculations have appeared in Quantum-Chemistry
literature over the past years. A brief account of these can be found in 
Ref.~\cite{bCasida}.

\section{Method and Results}
\label{MethodResults}
We used a real-space, real-time approach to solve the TD Kohn-Sham equations
for \({\rm Na}_2\), \({\rm Na}_4\), silane (\({\rm SiH}_4\))
and disilane (\({\rm Si}_2{\rm H}_6\)).
This time-evolution method is adequate to be combined with
real-space calculations of the ground-state by means of the finite-difference
pseudopotential method\cite{bmethod,yabana},
or with adaptative coordinates\cite{bGygi}.
In both cases, a real-space discretization of the kinetic energy
operator leads to sparse Hamiltonian matrices which do not need
to be stored in memory, and are easy to handle. Furthermore, in order to
propagate the states in time, we do not need to compute the complicated
\(f_{\rm xc}\) kernel, but only the much simpler \(v_{\rm xc}\) potential.
In this way, we are able to do time-dependent EXX calculations {\it without}
having to perform Slater's approximation \cite{bTDOEP,bSlater}.

The space was uniformly discretized, with mesh-spacing ranging from 
\(0.4\){\AA} for the sodium clusters to \(0.25\){\AA}  for the silicon ones,
and the points were contained inside a sphere with 
radius \(10\){\AA} for sodium and \(7\){\AA} for silicon.
The ionic potential was modeled by a soft-core Troullier-Martins
pseudopotential\cite{bTrouillerMartins}.
By using these settings, a convergence of better than \(0.1\)~eV was
reached for the eigenvalues.
The time-evolution was performed with a modified
Krank-Nicholson rule, with a time-step of 0.0025~$\hbar$/eV,
for a total evolution time, $T_{max}$,
of 50~$\hbar$/eV for the sodium clusters and 25~$\hbar$/eV for
the silicon clusters. With this parameters we can resolve excitation energies
within $\Delta E=2\pi\hbar/T_{max}$ that corresponds to tenths of eV.
We added absorbing boundary conditions
to improve the quality of the spectrum above the ionization
threshold (avoiding artificial formation of standing waves in
the spherical box used to confine the cluster).

To calculate the dipole strength function (which is simply
proportional to the absorption cross-section), we prepare
our system in the ground-state, and then excite it with a delta
electric field, \(E_0 \delta(t)\). The dipole strength can
then be related to the imaginary part of the dynamical
polarizability:
\be
  S(\omega) = \frac{4 \pi m_e}{h^2}\:\omega \Im\:\alpha(\omega)
  ,
\ee
where \(h\) is Planck's constant, \(m_e\) the electron's mass, and
the dynamical polarizability,
\be
  \alpha(\omega) = - \frac{2}{E_0} \int\! dr\: z \:\delta n(\br,\omega)
  .
\ee
In the last expression, \(\delta n(\br,\omega)\) stands for the
Fourier transform of \(n(\br,t) - n(\br,t=0)\).
The calculation of the dipole strength involves two important
approximations: The choice of the static xc potential used to obtain
the ground-state, and the time-dependent xc potential 
(directly related to the \(f_{\rm xc}\) kernel) used to propagate
the state. The relative importance of both is still fairly obscure.
Petersilka {\rm et al.} linear-response results\cite{bPeteretal} 
for the Helium and Beryllium atoms indicates that the choice of the
\(f_{\rm xc}\) kernel is less dramatic than the choice of a good
static \(v_{\rm xc}\) potential. However, this is no longer true
for the lowest excitation energies of Beryllium and if we look
at the singlet-triplet splittings, where the effects
arising from \(v_{\rm xc}\) cancel. The fairly accurate results of Lithium and
Beryllium within the LDA were related to the quality of the ground
state calculation for few electron systems as light atoms and the $H_2$ molecule\cite{pAry}.

By performing the time evolution, in the limit \(t\rightarrow\infty\),
we have all the information about the linear response of the system,
and should therefore be able to study the xc kernel {\it without} having
to perform explicitly the functional derivative of the xc potential.
Thus, we can study the effect of different \(f_{\rm xc}\)
kernels for a fixed \(v_{\rm xc}\) potential in the time-evolution
method. To address this question, we propose the following method:
First, we obtain the ground state of the system with the static
functional \(v_{\rm xc}^{(1)}\), and then perform the time evolution
with the xc potential:
\be
  \label{evxcfxc}
  v_{\rm xc}(\br,t) = v_{\rm xc}^{(1)}(\br,t=0) + \left[v_{\rm xc}^{(2)}(\br,t)
        - v_{\rm xc}^{(2)}(\br,t=0)\right] ,
\ee
where \(v_{\rm xc}^{(2)}\) is some other xc functional.
In this way, we probe the importance of the \(f_{\rm xc}^{(2)}\)
kernel, without having to spend more computational time in calculating
this rather complex quantity. This method will be used for silane,
for a varied selection of \(v_{\rm xc}^{(1)}\) and \(v_{\rm xc}^{(2)}\) to
quantify the dependence of the dynamical spectrum with the xc-potential and kernel.

In the following, PZ stands for LDA\cite{bKohnSham}, with 
\(\epsilon^{\rm hom}_{\rm c}\) taken from Monte-Carlo calculations
\cite{bCeperleyAlder}
and then parameterized by Perdew and Zunger \cite{bPerdewZunger}; PBE is the 
Perdew, Becke and Ernzerhof GGA functional \cite{bPBE}; LB94 represents
the van~Leeuwen and Baerends potential \cite{bLB94}; SIC is the SIC-LDA functional
\cite{bPerdewZunger}; and, finally, EXX stands for exact-exchange 
\cite{bTDOEP}. SIC and EXX are both treated within the KLI 
approximation \cite{bKLI}. ``exp'' will be used to denote
experimental values.
All energies are measured in eV unless otherwise stated.

\subsection{Sodium}


In Table~\ref{NaT1} we show the ionization potentials (IP),
obtained from the highest occupied KS eigenvalue, and
HOMO-LUMO gap for the two sodium clusters.
The IPs obtained either by using the LDA or the GGA
functionals are much smaller than the experimental
value. This is a well known problem related to the
wrong asymptotic behavior of the xc potentials. The
situation does improve enormously if we correct for
this deficiency, as can be seen from the LB94, SIC and EXX values.
Although the calculated IPs differ by more than \(2\)~eV,
the HOMO-LUMO gap stays essentially constant, {\it i.e}, correcting
for the asymptotic part of the exchange potential amounts to a constant
shift in the eigenvalues, and not to an opening of the gap.

All the calculated optical spectra for Na\(_2\) are quite similar (see Fig.~\ref{Fig1Na2}),
regardless of the xc potential used, and exhibit three clear
peaks in the 2-5~eV range (the third peak in the LB94 curve
is almost completely smeared out). They all compare
quite well with experiment, although the DFT peaks are all
shifted towards higher energies. This shift can be understood in terms of
the competition between the Coulomb repulsion in the electronic kernel and the 
electron/hole attraction from the xc part of the response (that is,
$\Delta\epsilon = \epsilon_c-\epsilon_v + <\varphi_v\varphi_c | \frac{1}{|r_1-r_2|}
| \varphi_v\varphi_c> + \Delta_{xc}$). 
In particular for a given valence-conduction transition \(v\rightarrow c\), the
exchange correlation kernel within the simple ALDA introduced only
a local and static {\it attractive} electron-hole interaction:
\bea
  \Delta_{xc}&=&\int\!\! d^3 r_1\; d^3 r_2 \: \varphi_v^*(\br_1) \varphi_c^*(\br_1) 
  \delta(\br_1-\br_2) \frac{\partial V_{xc}(\br_1)}{\partial n}
  \varphi_v(\br_2) \varphi_c(\br_2)
  \nonumber \\
  &=& \int\!\! d^3 r \: \varphi_v^*(\br) \varphi_v(\br) 
  \frac{\partial V_{xc}(\br)}{\partial n}
  \varphi^*_c(\br)\varphi_c(\br)
  .
\eea
This expression is more complicated
for the other kernels, but it is clear that they do introduce an effective
attractive interaction that it is not complete. We might have to recall that 
we have neglected dynamical effects  and they may need to be included in the xc kernel 
to recover this minor effect.
Also, we should keep in mind that temperature effects (vibrational motion) of the 
molecule may introduce a broadening of the spectrum as well as a 
shift of the peaks to lower frequencies \cite{bprlsi,pacheco}.
The functional which yields the best results for the dimer is, by a small margin, the EXX,
while the strongest depart from experiment is the LB94 curve.
We note that the 3rd peak in the LDA and PBE spectra lies in the continuum
of states (above the ionization threshold). Thus it is more a resonance
than a well defined bound transition as observed in experiments\cite{bNa2exp}.
This deficiency is observed in all the clusters studied in the present work 
and it is, once more, mainly due to the wrong 
description of the asymptotic part of the exchange potential.

For Na\(_4\) (Fig.~\ref{Fig1Na4}) all DFT calculations
yield very similar spectra (and similar to the many-body results
based on a $GW$ quasiparticle calculation and including the
electron-hole interaction through the solution of the Bethe-Salpeter
equation\cite{bNaMB}).
The spectra consist of three well separated peaks in the
1.5-3.5~eV range, and a broader feature at around 4.5~eV.
The comparison with the experimental values is quite good,
although the peaks appear, once more, at higher energies.
The best results were obtained using the LDA, the GGA and the EXX, while the LB94
spectrum showed the strongest deviation from experiment.
Again, the error is larger for the high energy peaks, where transitions to
conduction states close to the ionization threshold are involved.
Due to the small overlap between these Rydberg-like states and the
low-lying states, exchange processes are not relevant and the 
polarization contribution to correlation, although being generally weak
becomes the dominant contribution to 
the renormalization of these single particle excitations.

Finally, in Table~\ref{NaT2}, we present the static polarizabilities
of Na\(_2\) and Na\(_4\), obtained either through a finite field 
(static) calculation,
or from the Fourier transform of the time-dependent density.
For Na\(_2\) all functionals perform equally well,
being the results smaller than the experimental value by around
\(10\%\). This is consistent with other DFT calculations that
give static polarizabilities in the range 33.1-38.2~\AA\(^3\)
for Na\(_2\) and 67.1-78.7~\AA\(^3\) for Na\(_4\)
(see Ref.~\cite{bNaKuemmel} and results cited therein).
In the case of Na\(_4\), the orbital functionals
SIC and the EXX yield slightly better static polarizabilities than
the LDA and the GGAs, but the results are still smaller than the
experimental value. We also rejoice that the two methods used to calculate
the static polarizability yield very similar results. The neglect of
correlation as well as temperature effects is responsible for the
obtained smaller polarizabilities values. We indicate that the simple
argument that a potential with the correct asymptotic will lead
to more localized charge and therefore lower polarizability does not
hold for these small systems. These results give support to previous studies on simple 
metallic jellium spheres\cite{jellium}.

\subsection{Silicon}

We have calculated the two simplest hydrogen terminated silicon
clusters: silane (SiH\(_4\)) and disilane (Si\(_2\)H\(_6\)). These
systems pose a much harder challenge than the simple sodium
clusters (where even jellium calculations within the LDA yield
reasonable results), not only because of the presence of the
p-electrons, but also due to the hydrogen, which is very hard
to describe by a reasonably soft-core pseudopotential. As expected,
the five xc functionals we tried yielded quite dissimilar results.

Our calculations for the IP and HOMO-LUMO gaps for silane and disilane 
are summarized in Table~\ref{SiT1}. Undoubtfully, EXX and LB94
gave the best IPs of all five functionals tested, almost at the
level of the much more involved GW with exciton effects or
Monte-Carlo calculations. SIC is slightly worse, and LDA and PBE yield,
as usual, completely irreal IP. Although the IP changes by more than
\(4\) eV going from the LDA to the EXX, the HOMO-LUMO gap increases
by just around \(0.5\) eV: The main difference is, once more, a nearly
rigid shift of the eigenvalues. However, this shift brings the relevant
single particle transitions below the ionization threshold.
The results are also consistent with the finding that EXX calculations provide larger
gaps for semiconductors\cite{exx-gap}.

We will discuss separately the silane and disilane spectra.
The silane spectrum (Fig.~\ref{Fig1SiH4}) consists of three peaks between 8 and 12~eV
(the two peaks derive from a Jahn-Teller splitting of the
triply degenerate \(\rm 2t_2\rightarrow4s\) transition\cite{bJTsplit}),
followed by a much broader feature at higher energies.
The curves obtained with the two traditional
functionals, LDA and PBE, are quite similar to each other,
and the onset for absorption is underestimated by around 1~eV. The underestimation
of the onset of absorption is a well known deficiency of LDA based functionals that is
even more dramatic in the case of infinite bulk systems where excitonic effects and
band-gap renormalization are not properly described by these simple functionals\cite{excitons}.
The SIC spectrum is unphysically shifted to lower energies (fact that could
be anticipated by looking at the low SIC HOMO-LUMO gap), and the first
peak is split. This second effect is an artifact of SIC, which,
in this implementation, spontaneously breaks the 3-fold degeneracy
of the HOMO state. LB94 and EXX behave quite well: the onset for
absorption is now correct, and the error in the
position of the first three peaks is reduced by a factor of 2
from the LDA/PBE results (see Table~\ref{SiT3}).
van Leeuwen's functional performs marginally
better in this case than EXX, which overestimates slightly the
excitation energies. The transitions close to the
ionization threshold give rise to a peak in the spectrum that in experiments
is usually broader than in the calculations, as new decaying channels are available that are not
included in the present calculations. Although
the functionals we used do not properly include the damping
of the excitations, finite lifetimes can be simulated
by convoluting the calculated spectrum with some Lorentzian function.

The disilane LDA, PBE, LB94 and SIC spectra are all very similar (see Fig.~\ref{Fig1Si2H6}) 
and consist of five peaks in the 7-12~eV interval, followed
by a broader feature at higher energies. All these curves suffer
from the same deficiency: The separation between the second and 
third peak is too small, so the second peak appears as a shoulder
of the third due to the limited resolution of the calculations
(in Table~\ref{SiT3} the second line for Si\(_2\)H\(_6\)
describes the position of this shoulder). The curve closest to
experiment was calculated with the EXX functional (note its
first peak at exactly the experimental value). SIC yielded
a quite unreasonable spectrum, consisting of broad peaks
at too low energies, and consolidated its position of the worst
of the five functionals tested (as what concerns the calculation
of optical absorption spectra, of course).

For these clusters we also provide the static polarizabilities in Table~\ref{SiT2}.
For silane we find, in agreement with the results for sodium
clusters, that all the methods lead to polarizabilities lower
than the experimental values. In spite of the proper asymptotics of the EXX functional,
LDA yields the best results, which indicates that a subtle error cancellation
is taking place, and that it is important to have some sort of correlation
in the functional.
This result seems to hold true for these small systems and deserves
a more detailed analysis.


\subsection{Xc potential {\it vs} xc kernel}

In order to study the sensitivity of the optical spectra
to the xc functional used to calculated the ground-state,
and to the time-dependent xc potential used to
propagate in time, we plot, in Fig.~\ref{Fig1vxcfxc}, the
spectrum of silane, for some choices of \(v_{\rm xc}^{(1)}\) and
\(v_{\rm xc}^{(2)}\) (see Eq. (\ref{evxcfxc})). In the first plot
we take ground states calculated with several xc functionals
(EXX, LB94 and LDA), and propagate these states with the LDA,
while in the second
we fix the ground state (calculated with the EXX functional),
and use three different potentials to propagate it
(LDA, LB94 and EXX). The results clearly indicate that the
optical spectrum of SiH\(_4\) is much more sensitive to
the choice of the starting state, than to the quality of
\(v_{\rm xc}^{(2)}\). This numerical evidence is in good agreement 
with the findings of Petersilka {\rm et al.}\cite{bPeteretal}.
From the second plot of Fig.~\ref{Fig1vxcfxc} we see that the ALDA
provides the strongest attractive \(f_{\rm xc}\),
shifting the transitions to lower energies by as much as 0.4~eV, the 
effect being stronger for the high energy transitions.
Nevertheless, the role of these functionals needs to be quantified 
for larger systems and, in particular, for extended systems where
the form of the initial potential is expected to be less important and where
correlations are supposed to be the dominant contribution to the excitation
spectra. This different behavior of finite and infinite systems poses a
theoretical problem when designing functionals as the two classes of
systems are mainly sensitive either to the spatial non locality or to the
frequency dependence of the xc functional. 

\section{Conclusion}

We presented a detailed study of the impact of different
xc potentials and kernels in describing the static and
dynamical properties of small clusters. The ionization threshold as well
as specific single-particle transitions are very sensitive to the specific
form of the xc potential used to obtain the ground state of
the system.
The static polarizabilities turn out to be rather close to experiment but
always smaller than the observed values.
Surprisingly, the polarizabilities calculated with the LDA or the GGA are
better than the ones calculated with LB94, SIC or EXX. In the case of the EXX,
this is related to the absence of dynamical correlations, while the LB94 and the SIC
functionals seem to break the subtle error cancellation between
exchange and correlation always present in the LDA and the GGA functionals.

For the optical spectra a different conclusion is reached. EXX and LB94
perform remarkably well, even if the high lying excitations
(below the ionization threshold) appear shifted to higher
energies. This error is traced back to the weak effective attractive 
xc kernel. It seems clear that the right asymptotics 
are important but not enough to get a more quantitative agreement with
experimental data. SIC performs badly, especially for the
silicon clusters.
The more traditional functionals, the LDA and
the GGA, showed their well known problems: for example, 
the onset for optical absorption is underestimated by
around 1.5~eV in SiH\(_4\).

Finally, our spectra calculated with different xc-functional for the ground state calculation
and for the time evolution (\(v_{\rm xc}^{(1)}\ \ne v_{\rm xc}^{(2)}\)),
suggest that the choice xc potential used to calculated the ground state
is more important than the choice of the xc kernel for the case of finite systems. 
This conclusion should not be valid when one is dealing with infinite periodic systems.

\section{Acknowledgments}
We benefit from discussions with E.~K.~U.~Gross,
G.~Bertsch, K.~Yabana, L.~Reining, and K.~Capelle. We acknowledge
support by DGES (PB98-0345), JCyL (VA28/99), RTN program
of the European Union NANOPHASE (contract HPRN-CT-2000-00167).
M.~A.~L.~M. would like to thank the kind
hospitality and encouraging discussions with F. Nogueira.
A.~R. acknowledges the support from the sabbatical program
Salvador de Madariaga of the Spanish MEC (PR2000-0335)
and the Ecole Polytechnique.
A.~C. acknowledges the support from the MEC under the
graduate fellowship program.
Part of the calculations present in this
Article were performed at the Centro de F\'{\i}sica Computacional
of the Universidade de Coimbra.

\newpage
\section*{Table Captions}

Table~\ref{NaT1}: Calculated ionization potentials, (IP),
obtained from the highest occupied KS eigenvalue, and HOMO-LUMO gaps
for Na\(_2\) and Na\(_4\) for different xc functionals compared to
available experimental data\cite{bNaIP}. All values are in eV.

Table~\ref{NaT2}: Static polarizabilities for the small sodium clusters estimated
using a finite electrical field (FF), or calculated from the  $\omega=0$
Fourier transform of the time-dependent dipole  moment
(FT). The results are
compared to experiments \cite{bNaPol} and other calculations \cite{bNaKuemmel}.
All values are in \AA\(^3\).

Table~\ref{SiT1}: Calculated ionization potentials (IP)
obtained from the highest occupied KS eigenvalue, and HOMO-LUMO gaps
for SiH\(_4\) and Si\(_2\)H\(_6\). The available experimental data
as well as results from many body
calculations are given for comparison. DMC stands for the diffusion quantum
Monte-Carlo calculations of Ref.~\cite{bSiDMC},
GW for the GW with exciton effects results (GW-Bethe-Salpeter equation) of 
Ref. \cite{Grossmanetal}, and finally, QMC for quantum Monte-Carlo
calculations \cite{Grossmanetal}. All values are in eV.

Table~\ref{SiT2}: Static dipolar 
polarizabilities for the hydrogenated silicon clusters obtained
from the $\omega=0$ Fourier transform of the time-dependent dipole moment.
All values are in \AA\(^3\).

Table~\ref{SiT3}: Lowest excitation energies for SiH\(_4\) and Si\(_2\)H\(_6\) obtained with
different xc-functionals compared to experiments and other calculations
based on a many body formalism.
The SIC results are absent due to the difficulty to identify the
peaks in the spectra. DMC stands for the diffusion quantum
Monte-Carlo calculations of Ref.~\cite{bSiDMC}, and the calculated
value of 9.47 should be compared with the average of the two first, Jahn-Teller
splitted, levels.
GW are the GW with exciton effects results (GW-Bethe-Salpeter equation) of 
Ref. \cite{Grossmanetal}, and finally, QMC for Quantum Monte-Carlo
calculations \cite{Grossmanetal}. All values are in eV.

\newpage
\section*{Figure Captions}

\begin{figure}
\caption{} 
\label{Fig1Na2}
\begin{center}
\includegraphics[scale=1]{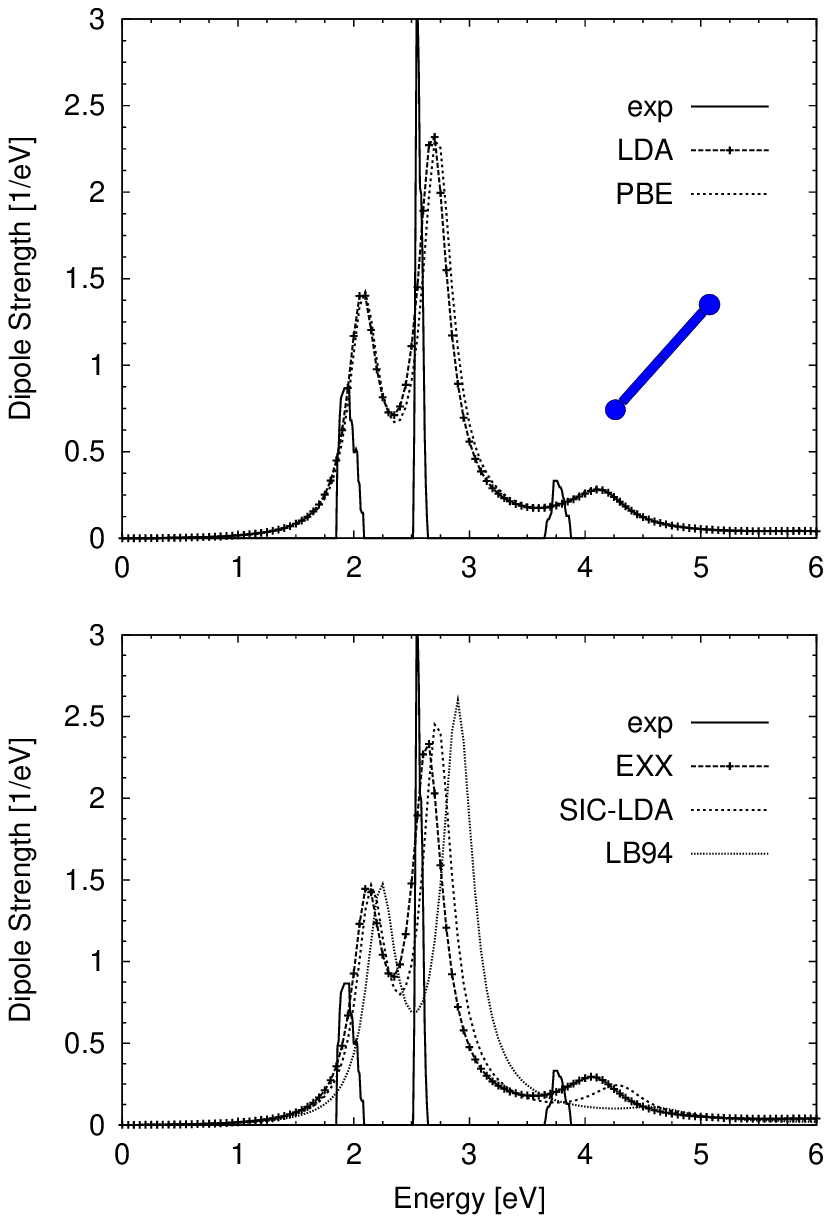}
\end{center}
\end{figure}

Fig.~\ref{Fig1Na2}: Averaged dipole strength for Na\(_2\), calculated for
several xc functionals. The ``exp'' curve is from
Ref.~\cite{bChelikowsky}, who adapted the experimental results of
Ref.~\cite{bNa2exp}. The experimental curve is plotted in arbitrary units.

\begin{figure}
\caption{}
\label{Fig1Na4}
\begin{center}
\includegraphics[scale=1]{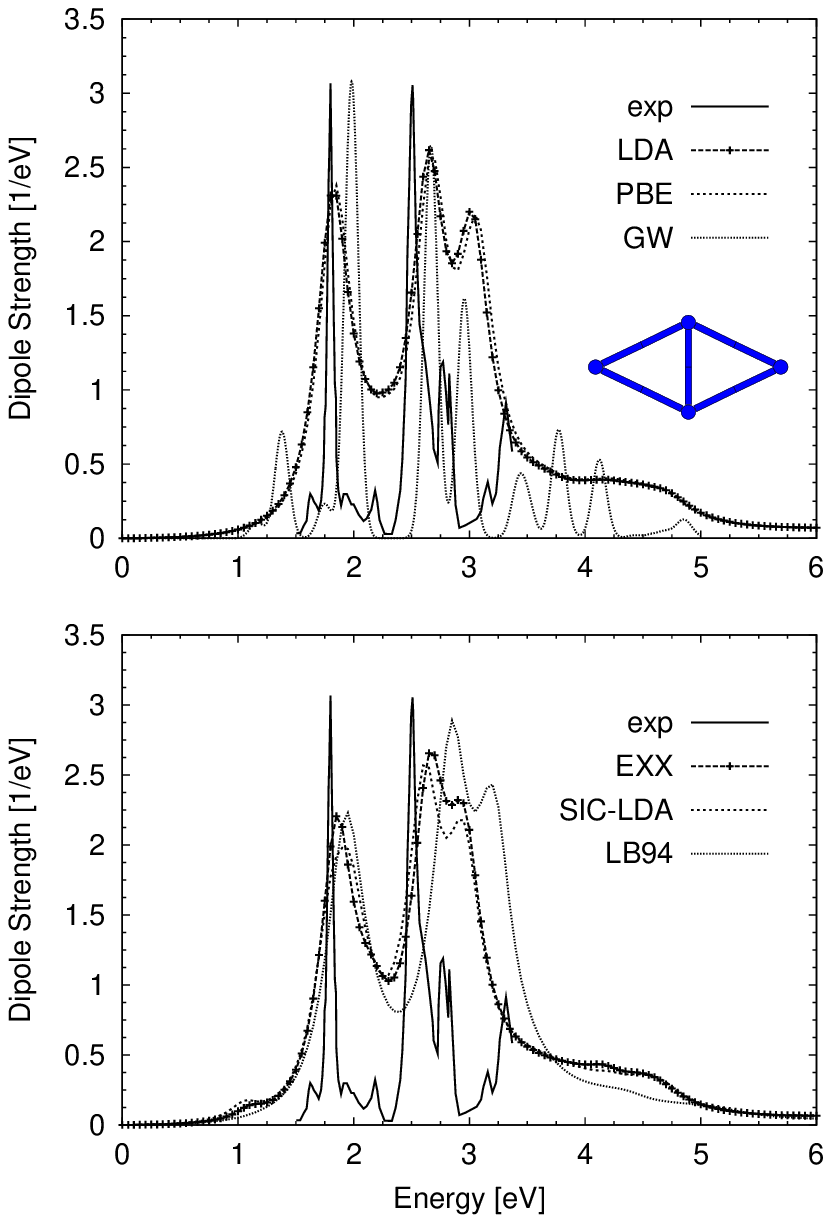}
\end{center}
\end{figure}

Fig.~\ref{Fig1Na4}:  Averaged dipole strength for Na\(_4\), calculated for
several xc functionals. ``exp'' stands for the
experimental photodepletion cross sections of Ref. \cite{bNa4exp}, 
while ``GW'' is the many-body calculation of
Ref.~\cite{bNaMB}, which includes self-energy and excitonic effects.
These two curves are shown in arbitrary units for the sake of comparison.

\begin{figure}
\caption{} 
\label{Fig1SiH4}
\begin{center}
\includegraphics[scale=1]{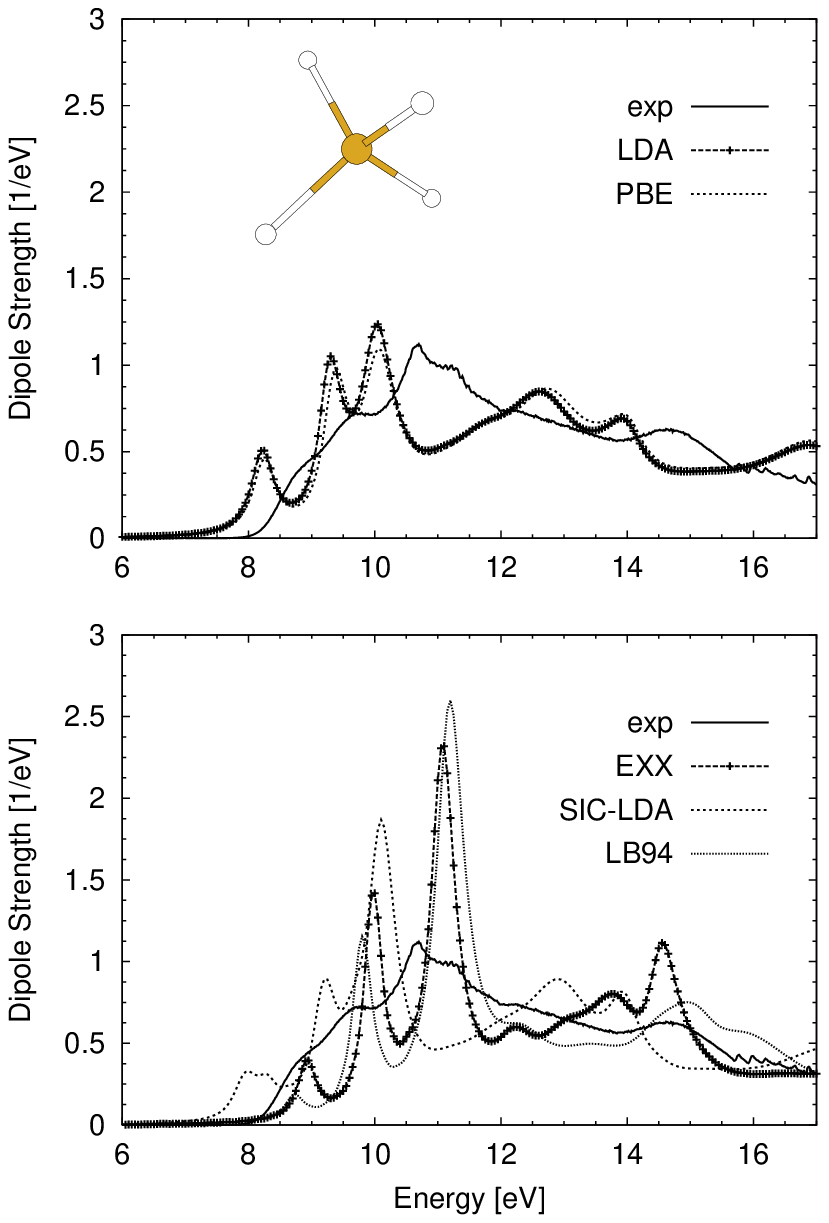}
\end{center}
\end{figure}

Fig.~\ref{Fig1SiH4}: Averaged dipole strength for SiH\(_4\), calculated for
several xc functionals. The ``exp'' curve is from
Ref.~\cite{bSiexp}.

\begin{figure}
\caption{}
\label{Fig1Si2H6}
\begin{center}
\includegraphics[scale=1]{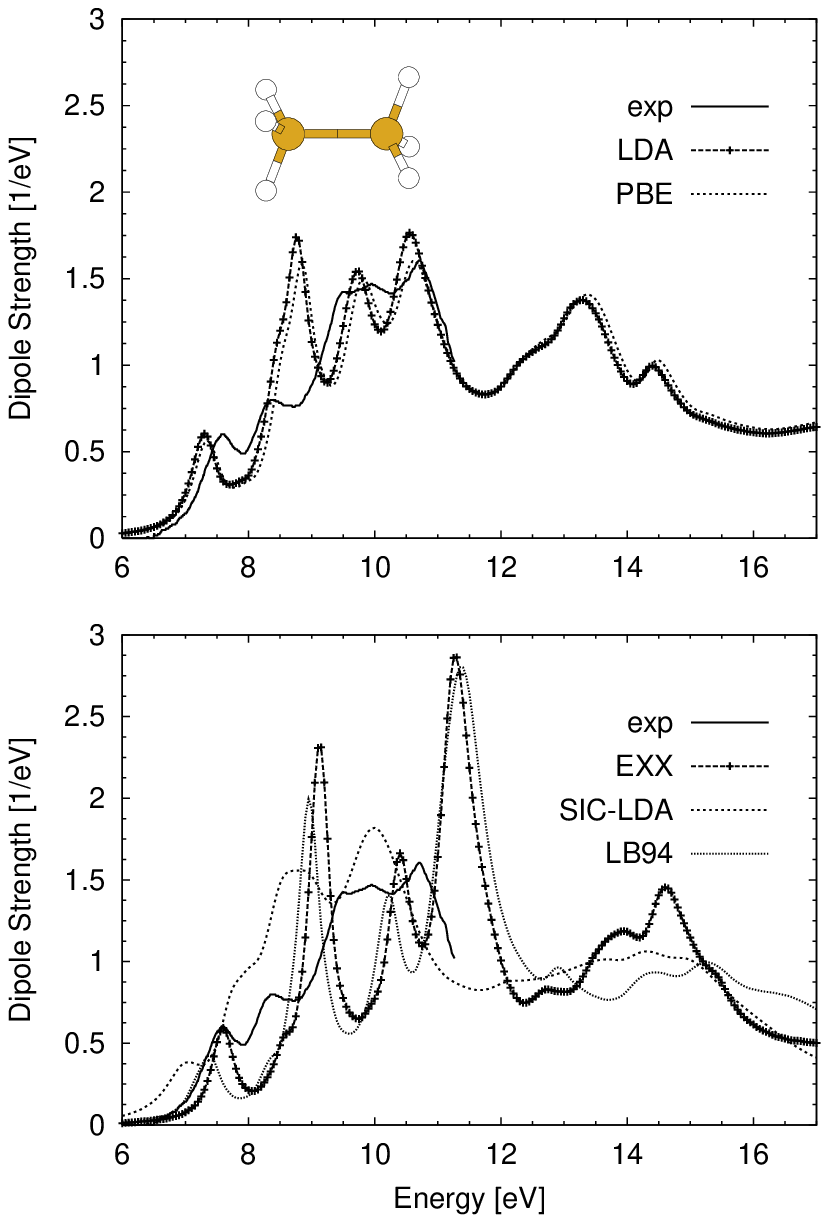}
\end{center}
\end{figure}

Fig.~\ref{Fig1Si2H6}: Averaged dipole strength for Si\(_2\)H\(_6\), calculated for
several xc functionals. The ``exp'' curve is from
Ref.~\cite{bSiexp}.

\begin{figure}
\caption{}
\label{Fig1vxcfxc}
\begin{center}
\includegraphics[scale=1]{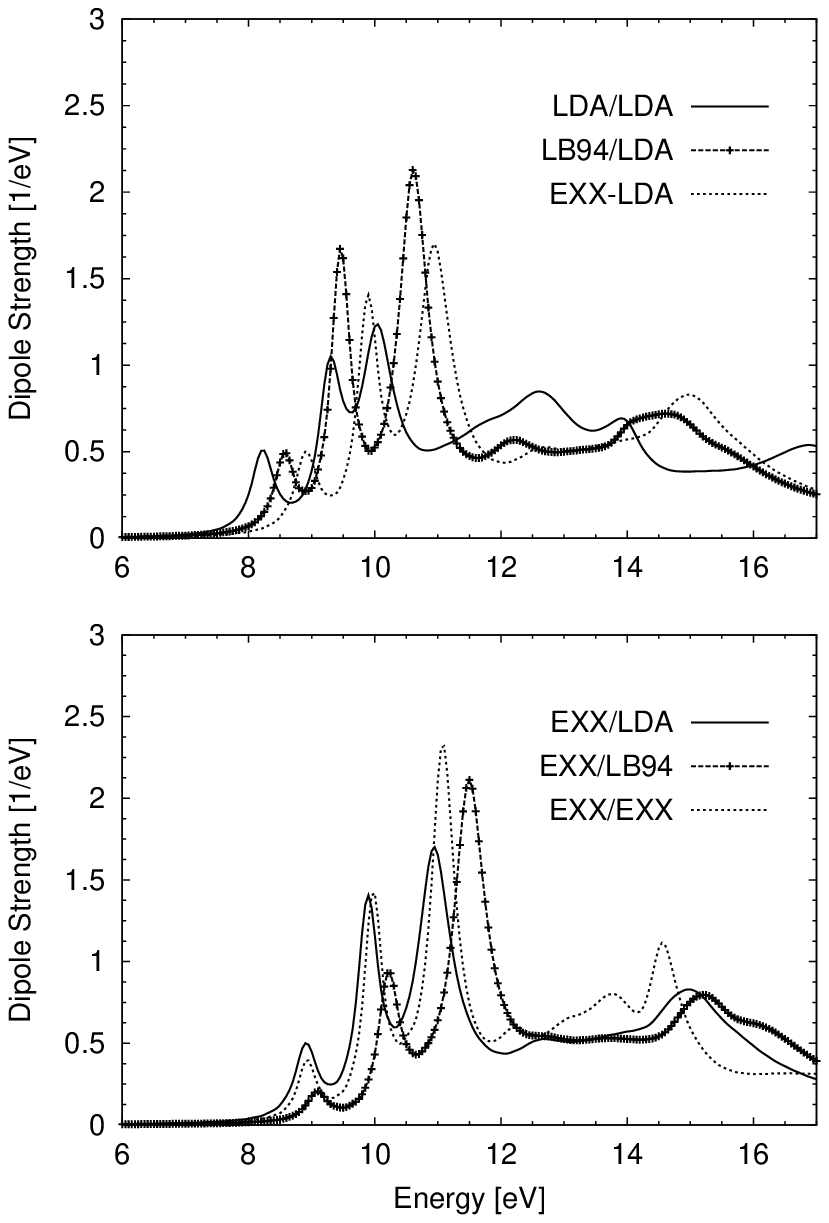}
\end{center}
\end{figure}

Fig.~\ref{Fig1vxcfxc}: Averaged dipole strength for SiH\(_4\), calculated
using different xc-functionals fro the calculation
of the ground state and for the time evolution
 (\(v_{\rm xc}^{(1)} \ne v_{\rm xc}^{(2)}\) ) (see
Eq. \ref{evxcfxc}). For each curve, the first functional
mentioned corresponds to \(v_{\rm xc}^{(1)}\) and
the second to \(v_{\rm xc}^{(2)}\)

\begin{table}
\caption{}
\label{NaT1}
\begin{tabular}{l|c|c|c|c|c|c}
                    & PZ   & PBE  & LB94 & SIC & EXX & exp \cite{bNaIP}\\ \hline \hline
Na\(_2\) IP         & 3.21 & 3.22 & 4.95 & 5.17 & 4.76 & 4.89 \\
Na\(_2\) HOMO-LUMO  & 1.36 & 1.36 & 1.25 & 1.51 & 1.45 &  -  \\ \hline
Na\(_4\) IP         & 2.75 & 2.75 & 4.44 & 3.85 & 3.88 & 4.27  \\
Na\(_4\) HOMO-LUMO  & 0.60 & 0.63 & 0.36 & 0.47 & 0.72 &  -  \\
\end{tabular}
\end{table}

\begin{table}
\caption{}
\label{NaT2}
\begin{tabular}{l|c|c|c|c|c|c|c}
              & PZ  & PBE & LB94 & SIC & EXX & Ref. \cite{bNaKuemmel} & exp \\ \hline \hline
Na\(_2\) FF   & 35.0& 34.3& 31.8 & 33.7& 34.9& 37.0& 39.3\\
Na\(_2\) FT   & 34.9& 34.1& 31.6 & 33.5& 34.7&  -  & 39.3\\ \hline
Na\(_4\) FF   & 76.7& 75.4& 73.4 & 80.0& 77.4& 78.7& 83.8\\
Na\(_4\) FT   & 75.9& 74.6& 71.7 & 78.6& 76.3&  -  & 83.8\\
\end{tabular}
\end{table}


\begin{table}
\caption{}
\label{SiT1}
\begin{tabular}{l|c|c|c|c|c|c|c|c|c}
              & PZ  & PBE & LB94 & SIC & EXX & 
  DMC \cite{bSiDMC}& GW \cite{Grossmanetal} & QMC \cite{Grossmanetal} & exp \\ \hline \hline
SiH\(_4\) IP         & 8.53& 8.55& 12.7 & 11.8& 13.1 & 12.88& 12.7  & 12.6 & 12.61 \cite{bSiIP}\\
SiH\(_4\) HOMO-LUMO  & 8.10& 8.12& 8.40 & 7.70& 8.77 &  -   & 13.0  & -    &  -    \\ \hline
Si\(_2\)H\(_6\) IP   & 7.40& 7.37& 11.2 & 9.95& 10.9 & 10.90&   -   & -    & 10.53-10.7 \cite{bSi2IP} \\
Si\(_2\)H\(_6\) HOMO-LUMO  & 6.76& 6.80& 6.58 & 5.98& 7.17 &  -   &   -   & -    &  -    \\
\end{tabular}
\end{table}

\begin{table}
\caption{}
\label{SiT2}
\begin{tabular}{l|c|c|c|c|c|c}
               & PZ  & PBE & LB94 & SIC & EXX & exp \cite{bCRC} \\ \hline \hline
SiH\(_4\)      & 5.11& 4.94& 4.66 & 4.71& 4.53& 5.44 \\ \hline
Si\(_2\)H\(_6\)& 9.87& 9.58& 9.27 & 10.0& 8.70&  -   \\
\end{tabular}
\end{table}

\begin{table}
\caption{}
\label{SiT3}
\begin{tabular}{l|c|c|c|c|c|c|c|c}
                   & PZ  & PBE & LB94 & EXX & 
 DMC \cite{bSiDMC} & GW \cite{Grossmanetal} & QMC \cite{Grossmanetal} &exp \cite{bSiexp} \\ \hline \hline
SiH\(_4\) 1st      & 8.23& 8.25& 8.75 & 8.93& 9.47&9.2 & 9.1& 8.8\\
SiH\(_4\) 2nd      & 9.30& 9.39& 9.80 & 9.98&  -  & -  & -  & 9.7\\
SiH\(_4\) 3rd      & 10.1& 10.1& 11.2 & 11.1&  -  & -  & -  & 10.7\\ \hline
\(\left<\left|\Delta\right|\right>\)   
                   & 0.52& 0.49& 0.22 & 0.27&  -  & -  & -  &  -  \\ \hline \hline
Si\(_2\)H\(_6\) 1st& 7.30& 7.35& 7.38 & 7.60&  -  & -  & -  & 7.6 \\
Si\(_2\)H\(_6\) 2nd& 8.50& 8.60& 8.42 & 8.63&  -  & -  & -  & 8.4 \\ \hline
\(\left<\left|\Delta\right|\right>\)   
                   & 0.20& 0.23& 0.12 & 0.12&  -  &  - & -  & -
\end{tabular}
\end{table}

\newpage

\newpage

\newpage

\newpage

\end{document}